\documentclass[conference]{IEEEtran}
\IEEEoverridecommandlockouts
\usepackage{cite}
\usepackage{amsmath,amssymb,amsfonts}
\usepackage{algorithmic}
\usepackage{graphicx}
\usepackage{textcomp}
\usepackage{xcolor}

\usepackage{multirow}
\usepackage{tabularx}
\usepackage{tabularray}

\DeclareMathOperator*{\minimize}{minimize }

\def\BibTeX{{\rm B\kern-.05em{\sc i\kern-.025em b}\kern-.08em
    T\kern-.1667em\lower.7ex\hbox{E}\kern-.125emX}}
\DeclareMathOperator*{\argmin}{arg\,min}

\begin{document}

\title{Learning Point Spread Function Invertibility Assessment for Image Deconvolution\\
\thanks{This work was supported by the VIE-UIS under the research project 3944. \\ Romario Gualdrón acknowledges the support of the IEEE SPS Scholarship. }
}

\author{{Romario Gualdrón-Hurtado$^{\dag}$, Roman Jacome$^{\ddag}$, Sergio Urrea$^{\ddag}$, Henry Arguello$^{\dag}$, Luis Gonzalez$^{\dag}$ }\\
$^{\dag}$ Department of Computer Science, $^{\ddag}$ Department of Electrical Engineering \\
\textit{Universidad Industrial de Santander} \\
Bucaramanga, Colombia \\
\{yesid2238324@correo., roman2162474@correo., sergio2228328@correo., henarfu@, gonzalez@\}uis.edu.co
}

\maketitle

\begin{abstract}

Deep-learning (DL)-based image deconvolution (ID) has exhibited remarkable recovery performance, surpassing traditional linear methods. However, unlike traditional ID approaches that rely on analytical properties of the point spread function (PSF) to achieve high recovery performance—such as specific spectrum properties or small conditional numbers in the convolution matrix—DL techniques lack quantifiable metrics for evaluating PSF suitability for DL-assisted recovery. Aiming to enhance deconvolution quality, we propose a metric that employs a non-linear approach to learn the invertibility of an arbitrary PSF using a neural network by mapping it to a unit impulse. A lower discrepancy between the mapped PSF and a unit impulse indicates a higher likelihood of successful inversion by a DL network. Our findings reveal that this metric correlates with high recovery performance in  DL and traditional methods, thereby serving as an effective regularizer in deconvolution tasks.  This approach reduces the computational complexity over conventional condition number assessments and is a differentiable process. These useful properties allow its application in designing diffractive optical elements through end-to-end (E2E) optimization, achieving invertible PSFs, and outperforming the E2E baseline framework.

\end{abstract}

\begin{IEEEkeywords}
Image deconvolution, computational imaging, diffractive optical element design.
\end{IEEEkeywords}

\section{Introduction}

{Image deconvolution (ID) is a fundamental process in image processing to recover a high-quality image from a blurred measurement. The challenges in image acquisition that require effective ID stem primarily from the degradation of images through various processes, including motion blur, defocus blur, and the inherent limitations of the implemented optical system. The convolution of the true scene often models these degradations with a point spread function (PSF), which represents the response of the imaging system to a point source or unit impulse~\cite{ gonzalez2009digital}. The goal is to invert the convolution process, a task complicated by the fact that the PSF may not be accurately known and that the inversion process is sensitive to noise present in the measurement~\cite{gonzalez2009digital}. Traditional ID techniques, such as the Wiener filter, have offered solutions based on analytical models of the PSF and the statistical properties of the noise. However, these methods face limitations, especially in cases of unknown noise distribution, and situations in which the PSF has zeros in the modulating transfer function~\cite{wiener1949extrapolation}. Further variational methods improve traditional filtering approaches by incorporating prior information via regularization functions such as total variation, or Tikhonov~\cite{wang2014recent}. Additionally, variational methods have recently included plug-and-play (PnP) priors in which denoisers are employed as proximal steps in the optimization algorithm~\cite{kamilov2023plug}. Convergence guarantees of these approaches rely heavily on the {condition number of the convolution matrix (built upon the PSF with a Toeplitz structure).} 

{Recent efforts have focused on solving the ID problem by using deep neural networks {(DNNs)}~\cite{xu2014deep,dong2020deep}, which has become the state-of-the-art in ID. However, there is no clear analysis of the properties that the {employed} PSF needs to achieve for  DNN-based methods to attain high recovery performance. 
To bridge this gap in measuring the PSF invertibility in DNN-based methods, this work proposes a metric that employs a non-linear approach to learn the invertibility of an arbitrary PSF using a neural network. The network is trained to map the PSF to a unitary impulse with a mean squared error (MSE) loss function. The main intuition of this approach is that, if for two arbitrary PSFs ($\mathbf{h}_1$ and $\mathbf{h}_2$), the network can achieve a lower mean squared error for $\mathbf{h}_1$, this means that deconvolution with $\mathbf{h}_1$ can obtain better performance than deconvolution with~$\mathbf{h}_2$.

The results have shown a significant correlation between the proposed metric and the performance in {traditional and DL-based} ID methods, such that in the cases where the metric decreases, the reconstruction MSE decreases, and vice-versa. The proposed metric also shows a reduction in the computational complexity in comparison with condition number computation. Due to the {reduced time and memory required} and given its differentiable nature, this metric is useful as a regularizer for designing diffractive optical elements {(DOEs)} in an end-to-end (E2E) optimization with a reconstruction network. We validate this for RGB imaging demonstrating an improved recovery performance by including the proposed metric as a regularization of the {PSF of the DOE} compared with the non-regularized E2E optimization.}

\section{Image Deconvolution Background}

Denote the target image $\mathbf{x} \in \mathbb{R}^n$, where $n$ is the image number of pixels and the optical system {point-spread function (PSF)} is denoted as $\mathbf{h}\in \mathbb{R}^{k}$, where {$k$ is the kernel size}. Then, the observations are formulated as  $\mathbf{y} = \mathbf{x}\ast\mathbf{h} + \boldsymbol{\epsilon} \in\mathbb{R}^m$ where  $m$ is the size of the convolved signal depending on {the} padding employed, $\boldsymbol{\epsilon}\in\mathbb{R}^m$ is an additive white Gaussian noise, and $\ast$ denotes the convolution operator. A multitude of studies have focused on recovering $\mathbf{x}$ from $\mathbf{y}$~\cite{chan2016plug,tobar2023gaussian,monga2021algorithm,dong2020deep,xu2014deep,kamilov2023plug,wang2014recent}. In these studies, the structure {and accuracy} of the PSF significantly affects the recovery performance. Here, we will describe some of the most relevant ID recovery approaches. 

\subsection{Wiener Deconvolution}

Wiener filtering (WF)~\cite{wiener1949extrapolation} is one of the most used tools in signal processing. The Wiener filter is the optimal filter for a minimum mean-squared error of a stationary random process. Consider the Fourier transform of the observations, PSF, target signal, and noise as $\widetilde{\mathbf{y}}=\mathbf{Fy}, \widetilde{\mathbf{h}}=\mathbf{Fh},\widetilde{\mathbf{x}}=\mathbf{Fx}$, and $\widetilde{\boldsymbol{\epsilon}}=\mathbf{F}\boldsymbol{\epsilon}$ respectively where $\mathbf{F}$ is the {discrete Fourier transform} matrix of an appropriate size. Via  WF, the recovered signal is obtained as $\widehat{\mathbf{x}} = \mathbf{F}^{-1}\mathbf{g}\odot \widetilde{\mathbf{x}}$, where $\odot$ is the Hadamard or element-wise product, $\mathbf{g}$ is the optimal WF which is computed as $\mathbf{g} = \frac{\mathbf{\widetilde{h}}^*}{\vert \mathbf{\widetilde{h}}\vert^2 + \sigma^2}$ where $*$ denotes the conjugate and $\sigma^2$ is the noise variance which is a {tunable hyperparameter}~\cite{wiener1949extrapolation}. The Wiener deconvolution fails to reconstruct the signal when the modulation transfer function of the PSF {(magnitude of $\widetilde{\mathbf{h}}$)}, has several zero-valued spatial frequency components~\cite{agrawal2009coded}. 

\subsection{Variational Methods}

Variational formulations for ID~\cite{likas2004variational}  introduce prior information on the target signals to reduce the ill-posedness of the inverse problem. Mainly, variational methods aim to solve
% the following optimization problem 
\vspace{-0.5em}
\begin{equation}
    \minimize_{\mathbf{x}} f(\mathbf{x}) + \rho g(\mathbf{x}),\label{eq:opt}
\end{equation}
% \vspace{-0.5em}
where $f(\mathbf{x})$ is known as the data fidelity term, which aims to promote consistency with the observation model, usually an $\ell_2$ norm is employed as $\Vert \mathbf{y}-\mathbf{H}\mathbf{x}\Vert_2^2$, where $\mathbf{H}\in \mathbb{R}^{m\times n}$ is the convolution matrix built upon the kernel $\mathbf{h}$, i.e., $\mathbf{H}_{i,i+j} = \mathbf{h}_j, i=0,\dots,m-1, j=0,\dots,k-1$, and $g(\mathbf{x})$ is a regularization term promoting the prior information of the scene, e.g., total variation $g(\mathbf{x}) = \Vert \sqrt{\vert\nabla_h\mathbf{x}\vert^2+\vert\nabla_v\mathbf{x}\vert^2} \Vert_1$, with $\nabla_h$ and $\nabla_v$ are horizontal and vertical derivatives on the image. 
The success of the recovery method depends significantly on the conditionality of the matrix $\mathbf{H}^T\mathbf{H}$, since most of the solvers for \eqref{eq:opt} require the gradient of $f(\mathbf{x})$. Here, we will use the definition of condition number i.e., {$\kappa(\mathbf{H}) =\frac{\lambda_{max}(\mathbf{H})}{\lambda_{min}(\mathbf{H})}$} where $\lambda_{max}$ ($\lambda_{min}$) is the largest (smallest) eigenvalue.  This metric is computationally expensive for large-scale {problems, i.e.} high-resolution images. 

\subsection{Deep Learning-based Deconvolution}

With the rise of deep learning {(DL)} to solve high-level computer tasks, data-driven ID methods have become the state of the art due to their ability to model non-linear interactions on the recovery and the use of diverse training datasets {~\cite{liu2020connecting}}. The goal of DL-based ID is to optimize a model $\mathcal{M}_\theta(\cdot): \mathbb{R}^{m}\rightarrow \mathbb{R}^{n}$ with trainable parameters $\theta$ employing a dataset, leading to 
\vspace{-0.5em}
\begin{equation}
    \theta^\star =  \argmin_\theta \mathbb{E}_{\mathbf{x}} \mathcal{L}(\mathbf{x}, \mathcal{M}_\theta(\mathbf{x}\ast\mathbf{h})),
\end{equation}

where $\mathcal{L}(\cdot,\cdot)$ denotes the reconstruction loss function. The network structure has been an extensive case of study, which varies from black-box deep neural networks~\cite{xu2014deep} to model-inspired {unrolling algorithm~\cite{monga2021algorithm} or {WF-inspired neural} networks such as Deep Wiener (DW)~\cite{dong2020deep}.}

\section{Learning Invertibility {Assessment}}

{Considering the limitations of traditional methods when working with PSFs that do not meet {the} expected properties and the high computational complexity of calculating the condition number of the PSF convolution matrix, the proposed approach is based on the following optimization problem to {learn to measure} the invertibility of a PSF in the ID problem}
\vspace{-0.5em}
\begin{equation}
    \beta^\star = \argmin_\beta \Vert\delta-\mathcal{N}_{\beta}(\mathbf{h})\Vert_2,\label{eq:metric}
\end{equation}
{with $\delta$ representing the unit impulse, and $\mathbf{\beta}$ being the trainable parameters of the neural network, it aims to measure the invertibility of the PSF by assessing the ability of the neural network $\mathcal{N}_{\beta}(\cdot)$ to map to a unit impulse PSF. {The key insight of this approach is that for two PSFs $\mathbf{h}_1$ and $\mathbf{h}_2$, the network is trained by solving \eqref{eq:metric} with standard gradient descent-based algorithms, leading to $\mathcal{N}_{\beta_1^\star}$ and $\mathcal{N}_{\beta_2^\star}$. Thus, if $ \Vert\delta-\mathcal{N}_{\beta_1^\star}(\mathbf{h}_1)\Vert_2 < \Vert\delta-\mathcal{N}_{\beta_2^\star}(\mathbf{h}_2)\Vert_2$, we state that ID with the PSF $\mathbf{h}_1$  can achieve better performance than with $\mathbf{h}_2$}. The network structure is the following}
\vspace{-0.5em}
\begin{align}
\hat{\delta} &= \mathcal{N}_{\beta}(\mathbf{h}) = \mathbf{W}_2(\psi(\mathbf{W}_{1} \mathbf{h} + \mathbf{b}_{1}))+\mathbf{b}_2,
\end{align}
{where $\psi$($\cdot$) is the Rectified Linear Unit activation function. $ \mathbf{W}_1 \in \mathbb{R}^{d\times k}$ is the {weights with $d$ hidden neurons and $\mathbf{b}_1\in\mathbb{R}^{d}$ is the corresponding bias vector of the first fully connected layer}, $\mathbf{W}_2 \in \mathbb{R}^{k\times d}$ and $\mathbf{b}_2 \in \mathbb{R}^{k}$ are the weights and biases of the output fully connected layer, respectively. Note that the parameters of the network are $\beta = \{\mathbf{W}_1,\mathbf{W}_2,\mathbf{b}_1,\mathbf{b}_2\}$.}

 It is well-known that additive noise in ID affects significantly the recovery performance, wherein the PSF structure can lead to better or worse deconvolution and denoising performance. Thus, {we also consider} measuring the PSF invertibility under the presence of additive {noise, as follows}
 \vspace{-0.5em}
\begin{equation}
    \beta^\star = \argmin_\beta \Vert\delta-\mathcal{N}_{\beta}(\mathbf{h}+\boldsymbol{\epsilon})\Vert_2,\vspace{-0.1cm}
\end{equation}
where $\boldsymbol{\epsilon} \in \mathbb{R}^{k}
$ is drawn from a Gaussian distribution $\boldsymbol{\epsilon} \sim \mathcal{N}(0,\sigma^2\mathbf{I})$ with variance $\sigma^2$.
 
% \RJcomment{It is missing the loss function, write it as equation 4.}

\section{Metric as Regularizer for DOE Design}

Currently, the design of diffractive optical elements (DOEs) for downstream tasks has opened new frontiers in computational imaging~\cite{arguello2022deep}.  While remarkable imaging performance has been achieved with the joint optimization of the {DOE heightmap} and reconstruction network, known as end-to-end (E2E) optimization there is no clear explanation of whether the imaging performance is due to the optimal lens design or due to the recovery network~\cite{arguello2022deep}. 

\subsection{Differentiable Physical modeling}

Mathematically, the light propagation through a diffractive lens is modeled with Fresnel propagation theory as follows: considering a wavefront before the diffractive lens {$u_o(x',y')=U(x',y')e^{\theta_0}$, with amplitude $U(x',y') $ and constant phase $\theta_0$, the wavefront after the diffractive lens is described as $u_1(x',y')=u_o(x',y')e^{i\psi(x',y')}$, where $\psi(x',y')=\frac{2\pi\Delta{\eta}}{\lambda}\phi (x',y') $}  is the phase added by the diffractive lens, where $\Delta{\eta}$ is the refraction index, the heighmap of the lens is parametrized via zernike polynomials such that $\phi (x',y')=\sum_{i=1}^L a_i Z_i$, where $Z_i$ are the zernike basis, $a_i$ the weight coefficients and $L$ is the number of employed polynomials. We want to find the optimal $a_i$ to generate the DOE. Then the encoded wavefront is propagated a distance $z$  to the sensor and the resulting wavefront can be modeled as {$u_2({x},{y},\lambda)= \frac{e^{jkz}}{jz\lambda}\iint u_1({x'},{y'},\lambda)e^{\frac{jk}{2z}((x'-x)^2+(y'-{y})^2)}\,d{x'} \,d{y'}$, $k=\frac{2\pi}{\lambda}$ is the wavenumber. Therefore, the PSF} acquired by the sensor is described as $h_\phi(x,y) \propto |\mathcal{F}\{U(x',y')e^{i\psi(x',y')}e^{i\frac{\pi}{\lambda z} (x'^2 + y'^2)} \}| $ by representing the Fresnel integral as a Fourier transform{~\cite{goodman2005introduction}.} {Finally,} the spectral image formation model can be represented as $g(x,y)= \int_{\lambda} {\gamma_{c}(\lambda) (I_\lambda * h_\phi)(x,y)} d\lambda $, where $\gamma_{c}(\lambda)$ is the wavelength sensitivity per channel $c$ of the sensor and $I_\lambda$ is the true scene in the wavelength $\lambda$. We can derive a discretized and vectorized model as 
$\mathbf{y} = \mathbf{H}_\phi\mathbf{x}$, where $\mathbf{H}_\phi$ is the sensing matrix parametrized with the heightmap $\phi$ which models the convolution and wavelength sensitivity operations.

\subsection{End to End Optimization}

Based on the physical modeling for rendering the {PSF of the DOE, E2E aims to optimize the DOE heightmap $\phi$ jointly} with a reconstruction network $\mathcal{M}_\theta$. The E2E optimization problem is formulated as 
\vspace{-0.5em}
\begin{equation}
    \{\phi^\star, \theta^\star\} = \argmin_{\phi, \theta} \mathbb{E}_{\mathbf{x}}\left[\mathcal{L} (\mathbf{x},\mathcal{M}_\theta(\mathbf{H}_\phi\mathbf{x}))\right].
\end{equation}

Note that updating the heightmap $\phi$ via gradient descent algorithms depends only on the loss function, in which a large recovery network $\mathcal{M}_\theta(\cdot)$ produces gradient {vanishing in the DOE optimizaion~\cite{jacome2023middle}}, leading to flawed DOE design. Thus, to incorporate the proposed invertibility metric, we devise the following bi-level optimization problem
\vspace{-0.5em}
\begin{align}
   \{\phi^\star, \theta^\star\} = & \argmin_{\phi, \theta} \mathbb{E}_{\mathbf{x}}\left[\mathcal{L}(\mathbf{x},\mathcal{M}_\theta(\mathbf{H}_\phi\mathbf{x}))\right] + \gamma \Vert \delta -\mathcal{N}_{\beta^\star}(\mathbf{h}_\phi)\Vert_2\nonumber\\ 
    &\text{subject to } \beta^\star = \argmin_\beta \Vert\delta-\mathcal{N}_{\beta}(\mathbf{h}_\phi)\Vert_2\label{eq:inner}
\end{align}

where $\gamma$ is a regularization parameter. {In this case, the proposed regularization allows for the consideration of the PSF invertibility in the E2E optimization. This ensures that the optimized heightmap $\phi^\star$ has increased invertibility, thus enhancing performance in the ID task. Also, note that the regularization is differentiable with respect to the heightmap since it is parametrized with the {DNN $\mathcal{N}_\beta(\cdot)$.}}
\section{Results}
We implemented the proposed metric in PyTorch~\cite{paszke2019pytorch}. As a common configuration in every experiment, we employed Adam optimizer~\cite{kingma2014adam}, with a learning rate of $1\times 10^{-3}$ for 1000 epochs, the size of the hidden layer was set to $d= 2048$. The dimension of the images was set to $128\times 128$ i.e., $n=128^2$, and the kernel size was set the same as the image size {$k=n$}. 
To measure the correlation between the proposed metric and the traditional method of assessing invertibility via the condition number, various input PSFs $\mathbf{h}$ with diverse characteristics were utilized to observe their behavior in different environments. 
\begin{figure}[!t]
    \centering
    \includegraphics[width=0.95\linewidth]{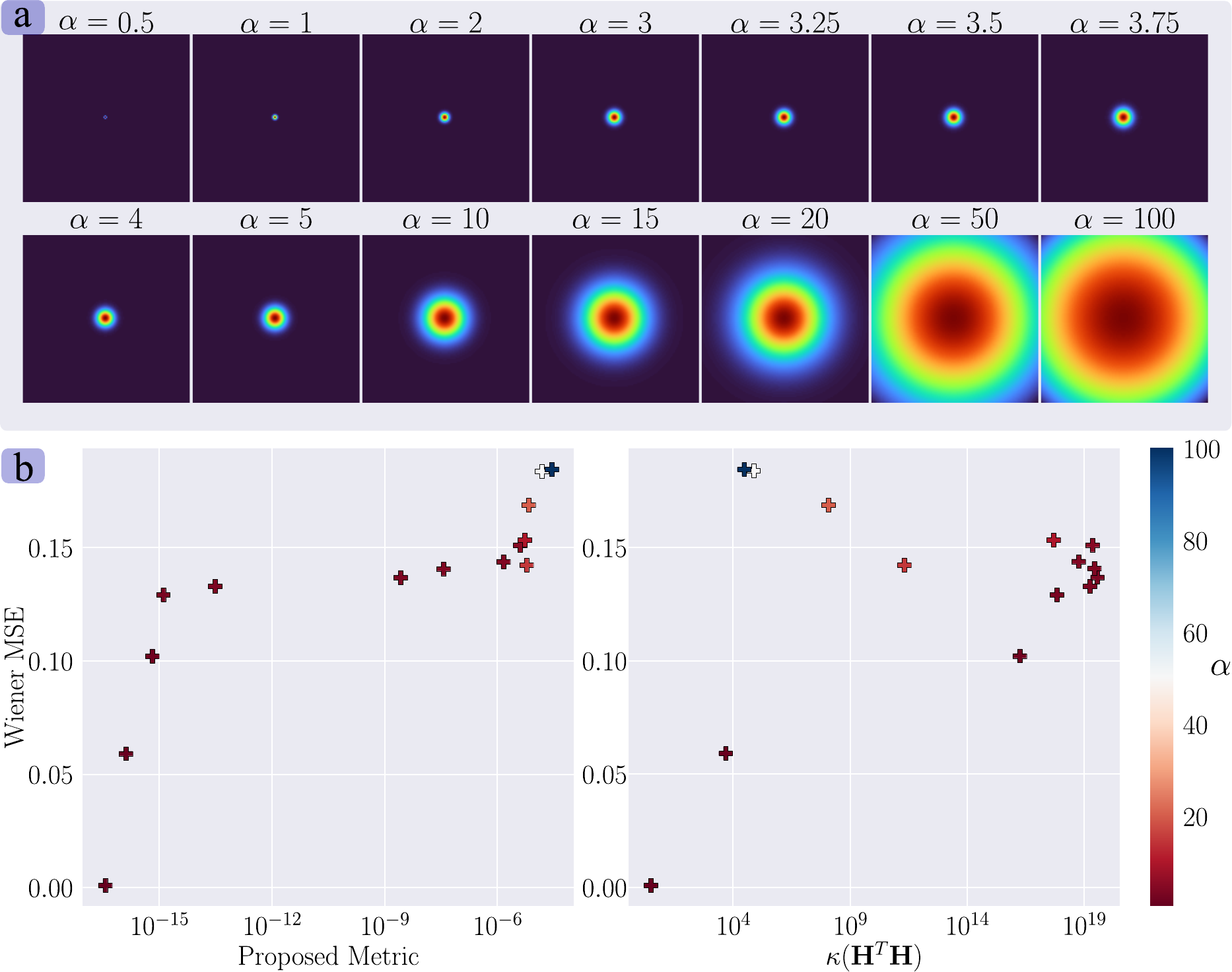}
    \caption{a) Employed Gaussian filters. b) Comparison of proposed vs. traditional invertibility assessment in Gaussian filters.}\vspace{-0.2cm}
    \label{fig:gaussians}\vspace{-0.2cm}
\end{figure}
\begin{figure}[!t]
    \centering
    \includegraphics[width=0.95\linewidth]{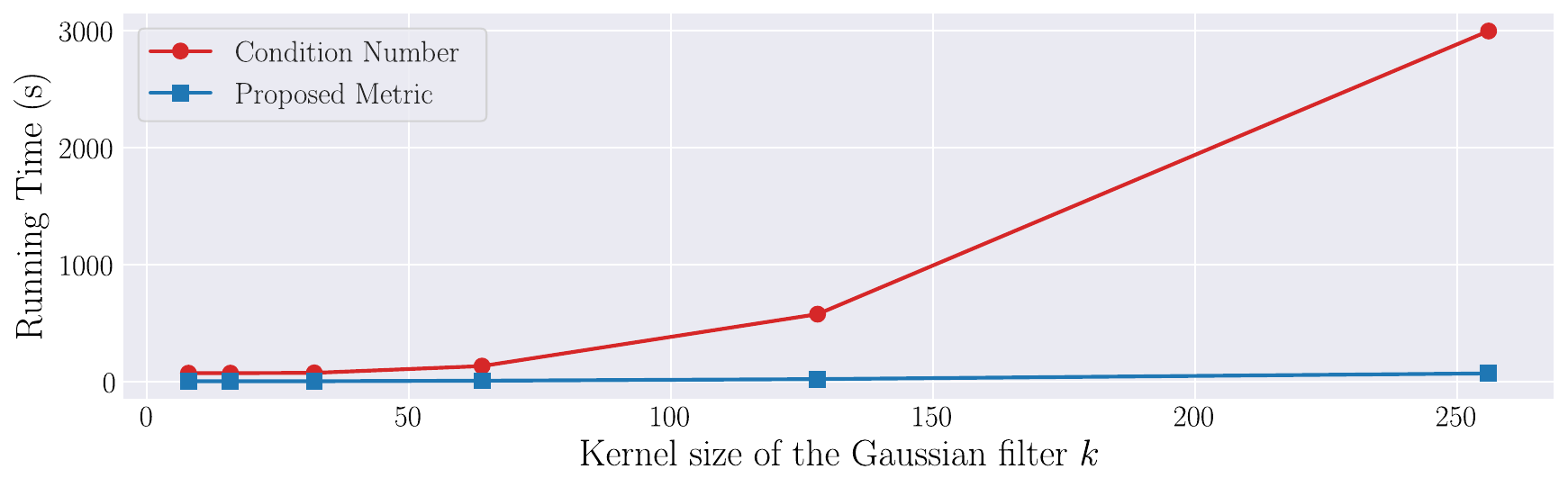}\vspace{-0.25cm}
    \caption{Running time comparison of proposed vs. traditional invertibility assessment in Gaussian filters with different kernel sizes.}\vspace{-0.5cm}
    \label{fig:time_gaussians}
\end{figure}

\textbf{Gaussian PSFs invertibility:}
% {Initially, in {Fig. \ref{fig:gaussians}}, an analysis is proposed using Gaussian PSFs with a variation in their standard deviation $\alpha$, where, as expressed in~\cite{tobar2023gaussian}, the difficulty of deconvolution increases proportionally to $\alpha$. Figure \ref{fig:gaussians} reveals a comparison between the proposed metric and the condition number of $\mathbf{H}^T \mathbf{H}$ to Wiener MSE, illustrating reconstruction quality. {Lower $\alpha$ correlate with reductions in both the proposed metric and Wiener MSE, suggesting enhanced reconstruction quality. In contrast, higher $\alpha$ increase the proposed metric, indicating greater inversion difficulty, which is directly associated with elevated Wiener MSE, denoting poorer reconstruction quality.} 
In {Fig. \ref{fig:gaussians}}, an analysis using Gaussian PSFs with varying standard deviation $\alpha$ is presented, as detailed in~\cite{tobar2023gaussian}, where larger $\alpha$ values complicate deconvolution. Figure \ref{fig:gaussians} compares the proposed metric to the Wiener MSE and the condition number of $\mathbf{H}^T \mathbf{H}$, showcasing that lower $\alpha$ improves both metrics, indicating better reconstruction quality, while higher $\alpha$ increases inversion difficulty.
This relationship shows a near-linear trend in extreme cases, highlighting the predictive accuracy of the metric for PSF invertibility. Conversely, the condition number of $\mathbf{H}^T \mathbf{H}$ exhibits irregular behavior diverging for {a higher $\alpha$}, leading to a lower condition number despite the worst reconstruction quality. The condition number fails to adequately measure the PSF invertibility in deconvolution tasks, unlike the proposed metric which achieves this with significantly less computational effort. To further validate this last claim, we analyze the running time of the condition number computation {and the proposed metric} in Fig. \ref{fig:time_gaussians} for different kernel size $k$. 
This shows that our method achieves almost constant running time while increasing $k$ whereas the condition number computation running time increases {at} a much higher rate for larger $k$.
}

\begin{table*}[!t] % Use table* for double-column tables in two-column documents
\centering
\caption{Comparison of Proposed vs. Traditional Invertibility Assessment for a Variety of PSFs.}\vspace{-0.2cm}
\label{tab:psfs}
\begin{tblr}{
  % colspec = {Q[c,86] Q[c,48] Q[c,m,85] Q[c,m,85] Q[c,m,85] Q[c,m,85] Q[c,63] Q[c,83] Q[c,m,85] Q[c,m,85] Q[c,m,85]},
  colspec = {Q[c,m,75] Q[c,m,75] Q[c,m,75] Q[c,m,75] Q[c,m,75] Q[c,m,75] Q[c,m,75] Q[c,m,75] Q[c,m,75] Q[c,m,75]},
  column{1} = {c},
  column{1-Z} = {c,m},
  cell{1}{3} = {c=2}{c},
  cell{1}{5} = {c=2}{c},
  cell{1}{7} = {c=2}{c},
  cell{1}{9} = {c=1}{c},
  vlines,
  hlines,
}
& \textbf{Impulse} & \textbf{Fresnel}~\cite{yeh2010analysis}    & & \textbf{Spiral}~\cite{jeon2019compact} & & \textbf{Motion blur}~\cite{dai2008motion}&   & \textbf{Privacy}~\cite{arguello2022optics}   \\

$\mathbf{h}$  &  \includegraphics[width=\linewidth]{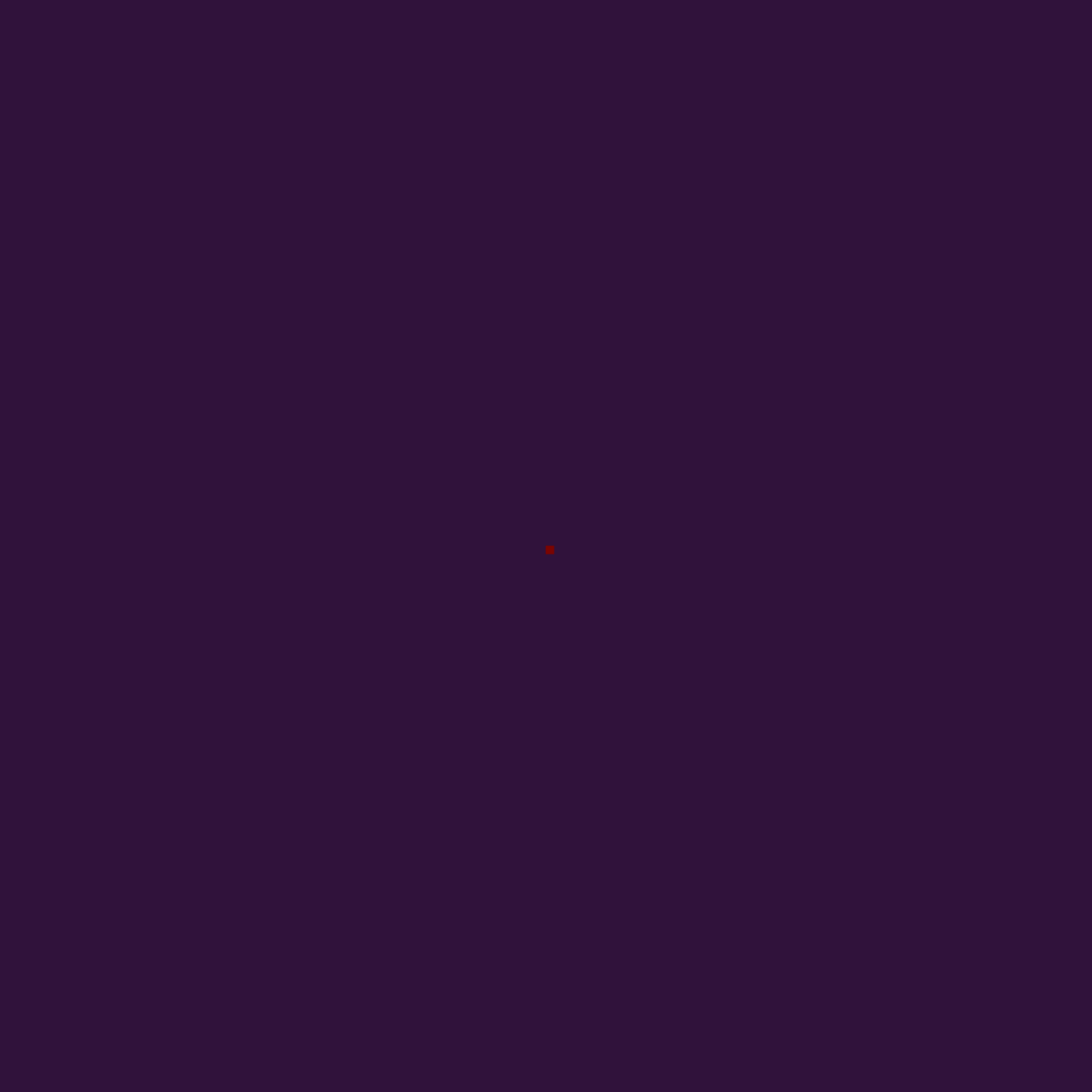}  & \includegraphics[width=\linewidth]{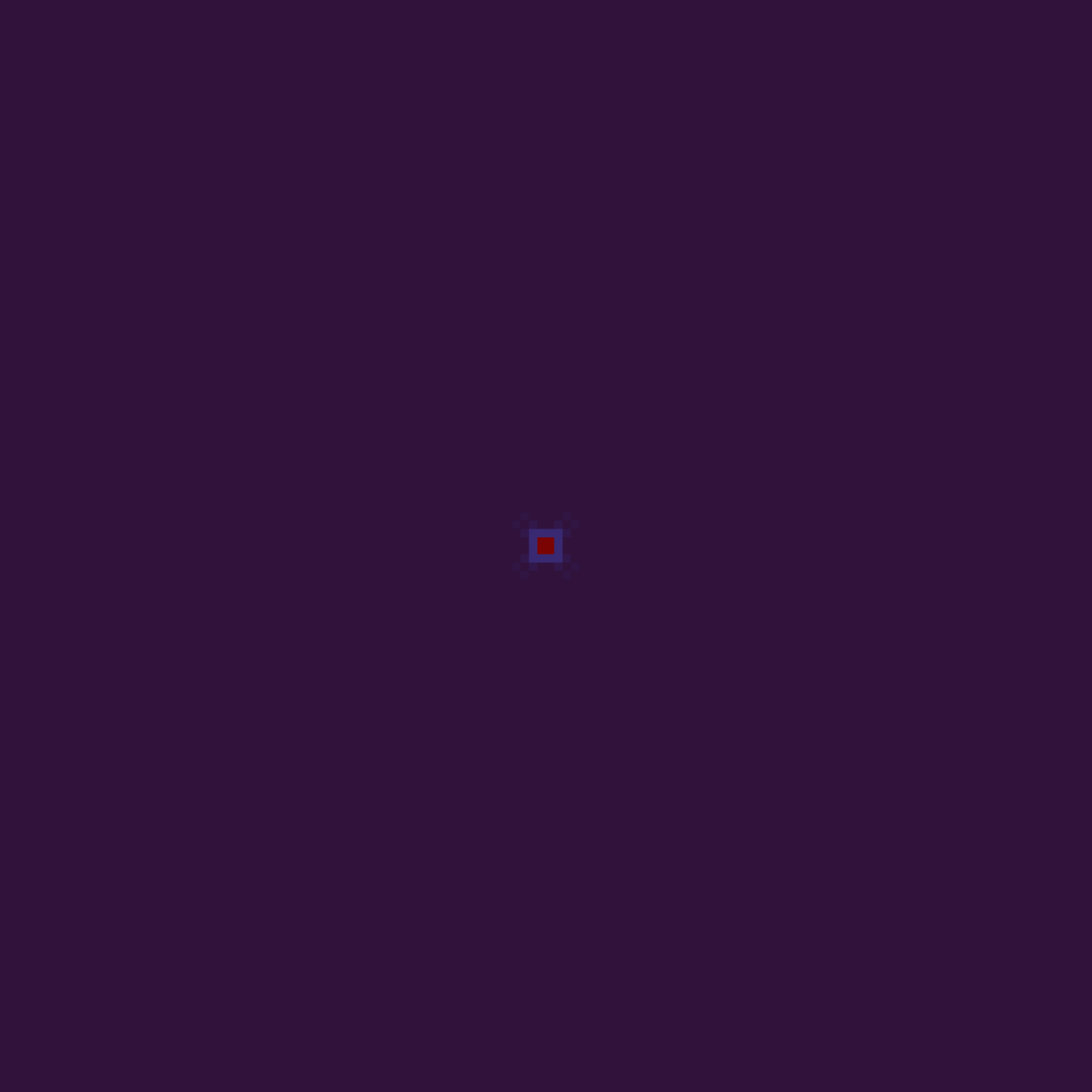}    & \includegraphics[width=\linewidth]{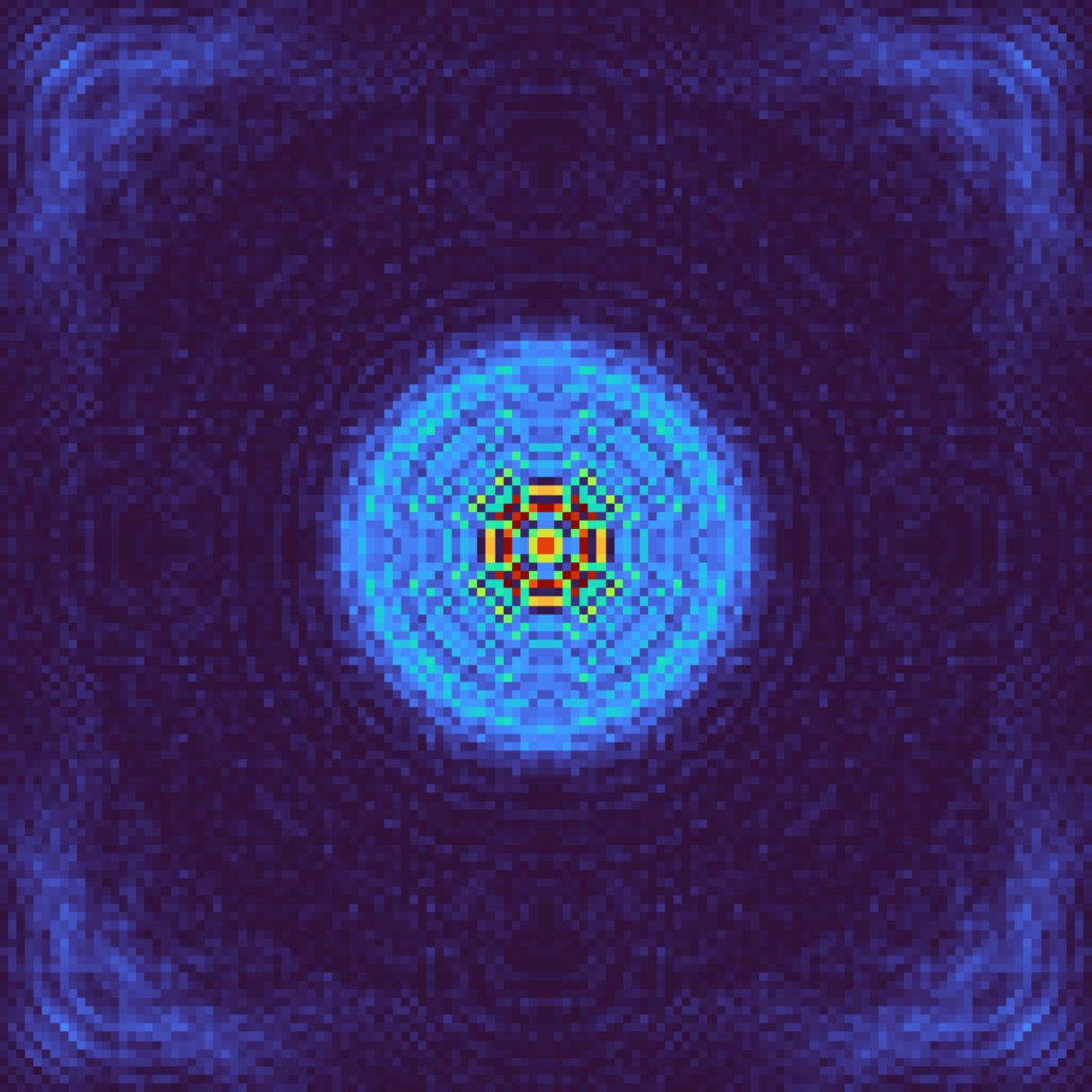}    & \includegraphics[width=\linewidth]{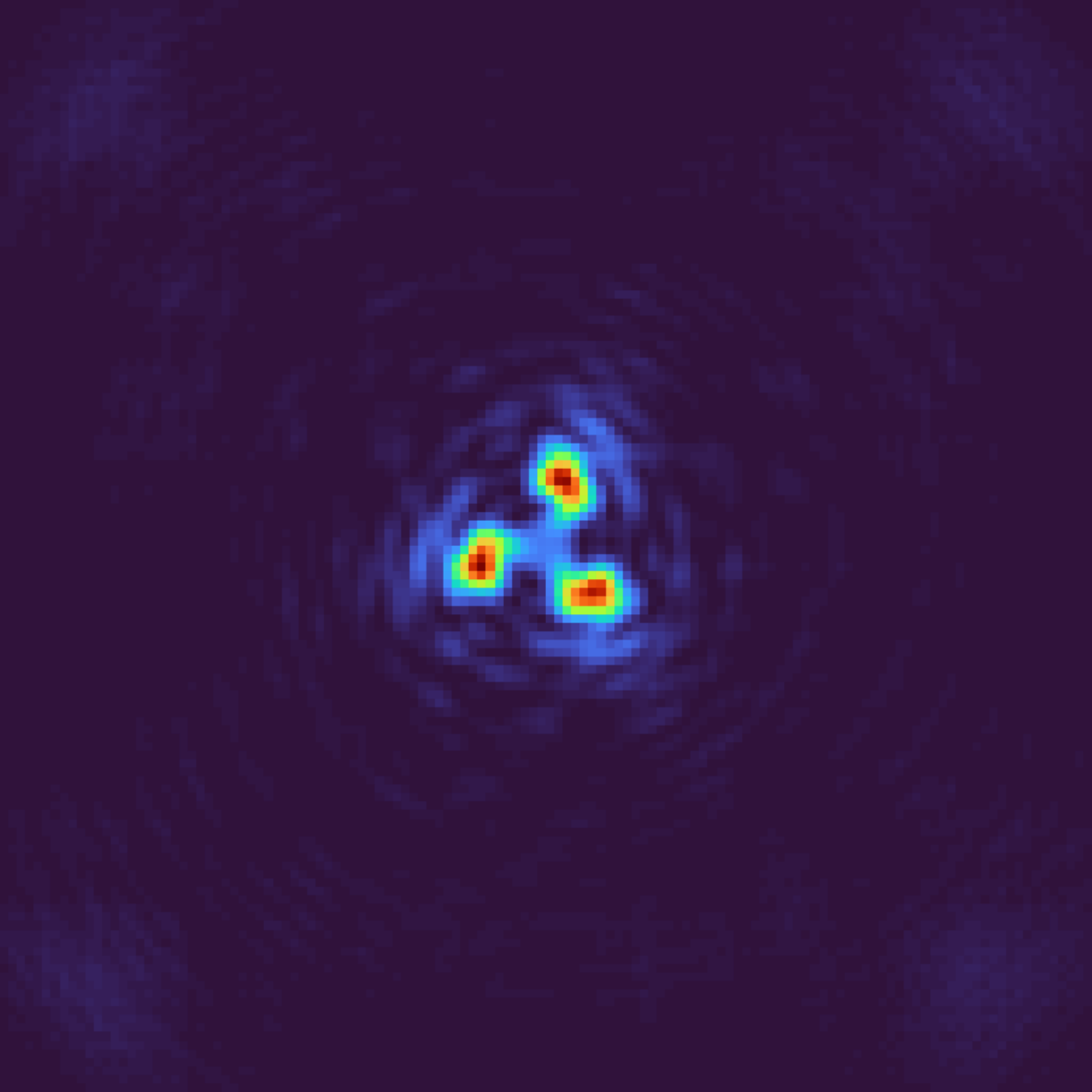}    & \includegraphics[width=\linewidth]{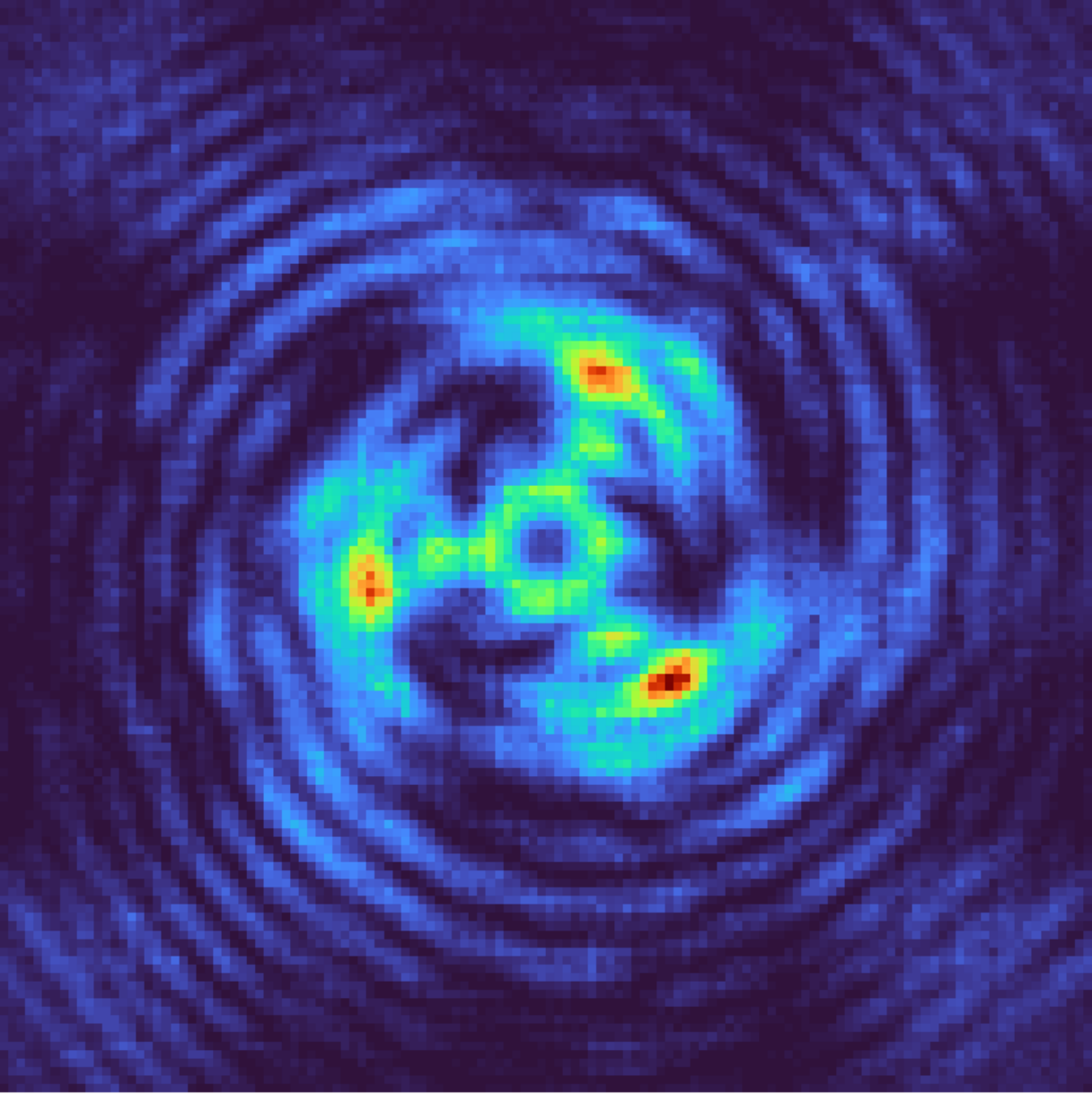}    & \includegraphics[width=\linewidth]{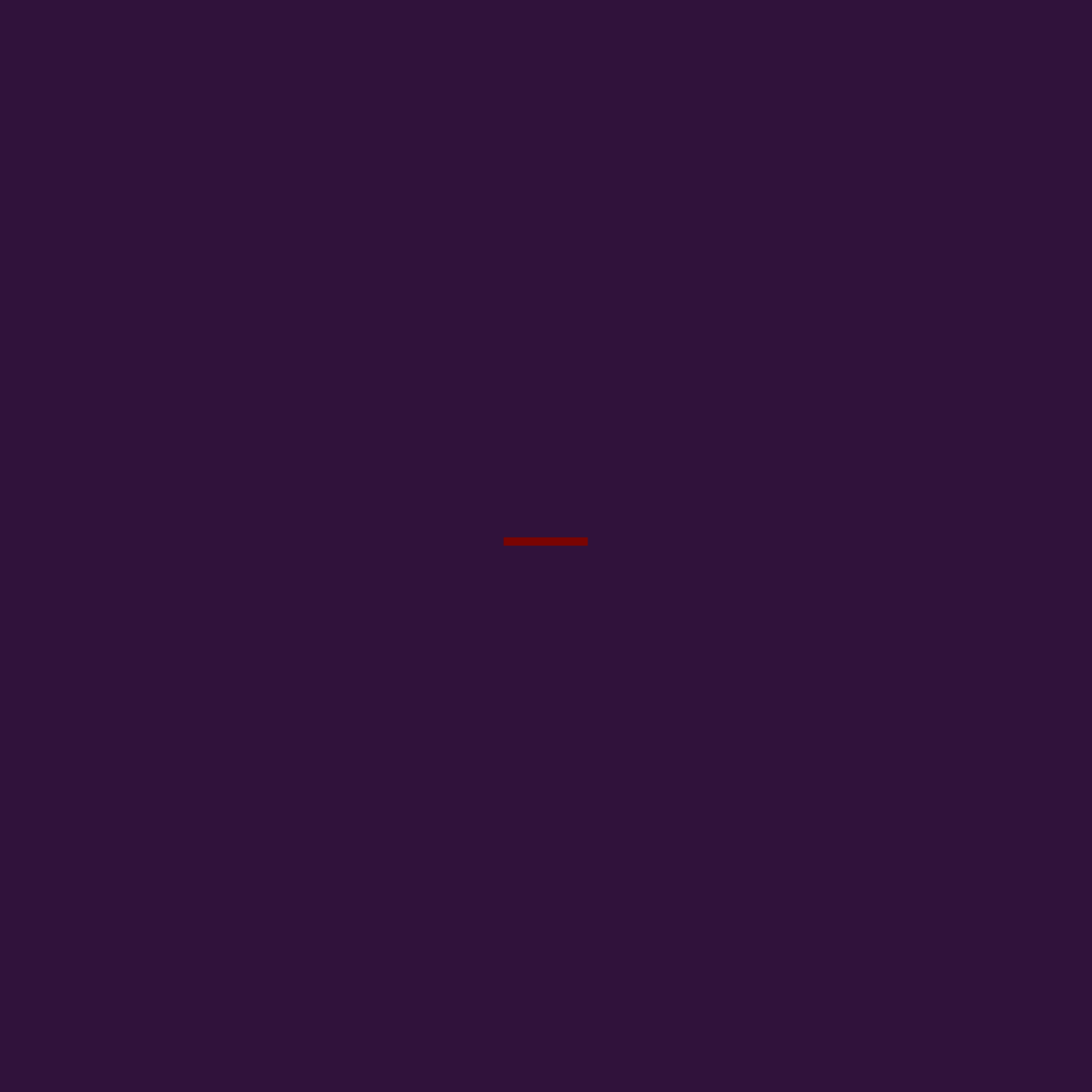}    & \includegraphics[width=\linewidth]{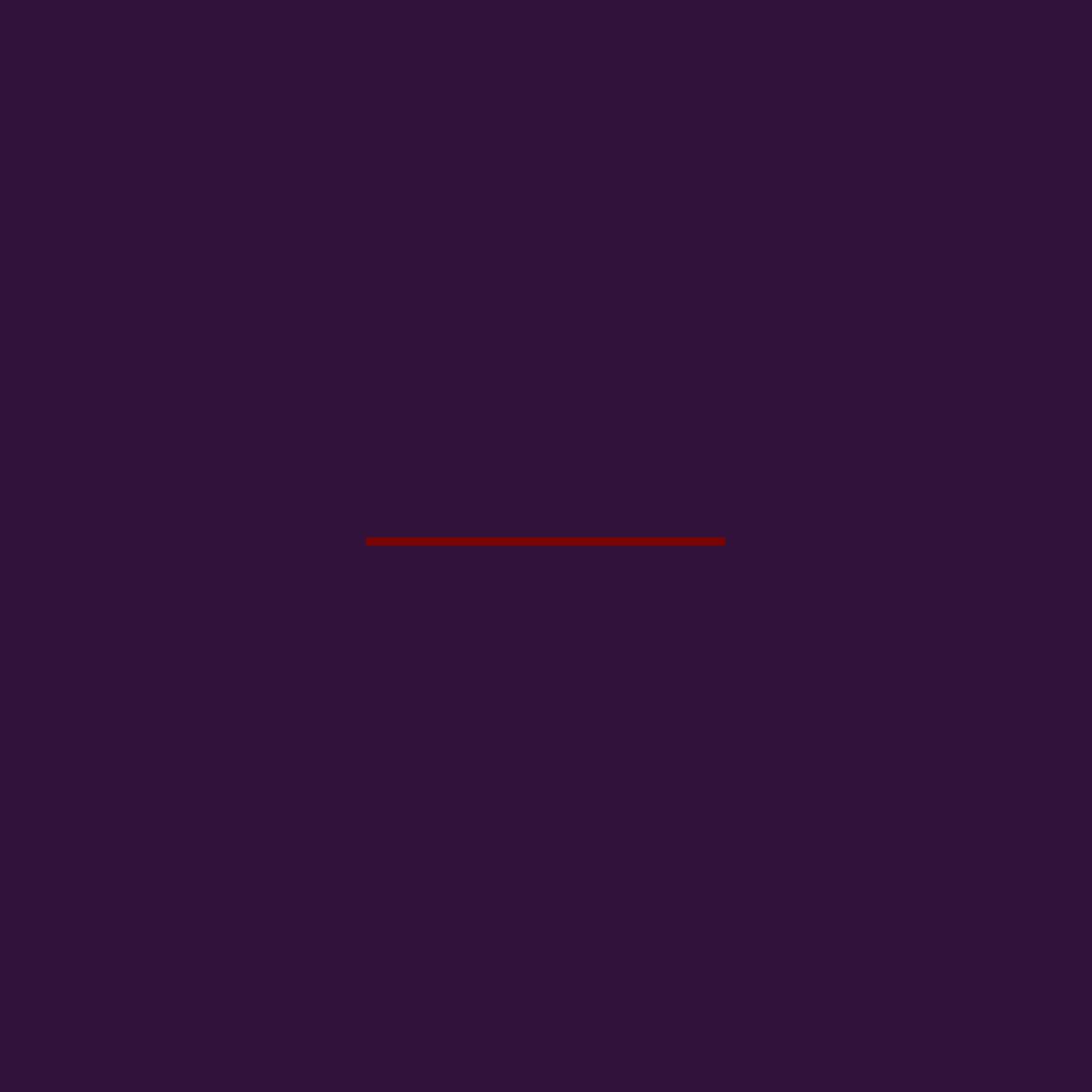} & \includegraphics[width=\linewidth]{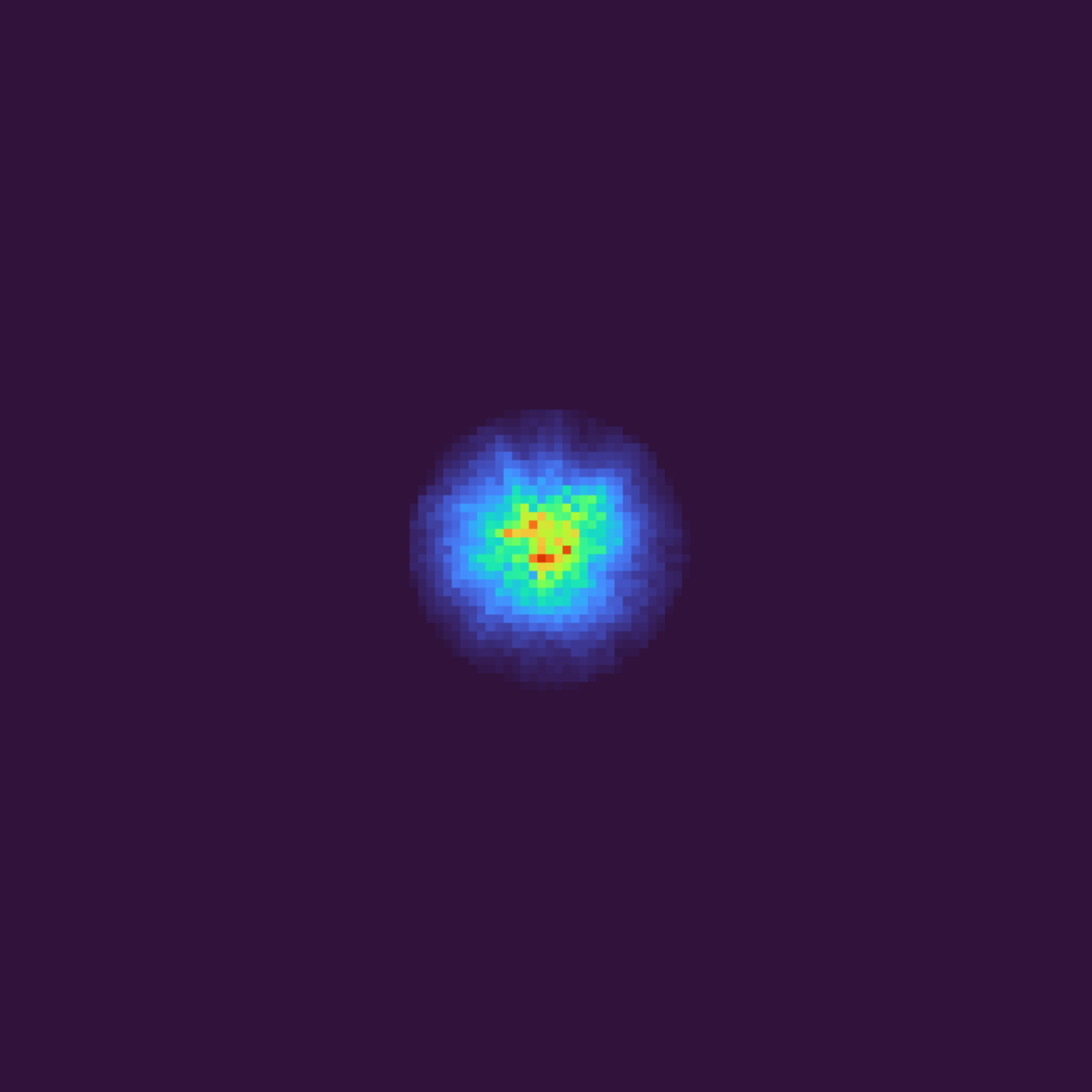} \\

% $\tilde{\kappa}(\mathbf{h})$ [PSNR] & $\infty$ & $174.19$ & $117.44$ & $132.16$ & $91.10$  & $152.43$ & $146.26$  & $54.39$  \\
Metric & $0$ & $2 \times 10^{-16}$ & $4.3 \times 10^{-6}$ & $4.1 \times 10^{-6}$ & $4.2 \times 10^{-6}$ & $4 \times 10^{-16}$  & $2 \times 10^{-15}$ & $3.6 \times 10^{-6}$   \\

$\kappa(\mathbf{H})$              & $1$ & $922.70$            & $916.86$            & $32676.5$           & $11556.3$           & $88.08$ & $11.58$ & $1406.1$ \\

% $\kappa_1(\mathbf{H}^T\mathbf{H})$  & $1$ & $4.2 \times 10^{6}$ & $1.0 \times 10^{7}$ & $4.9 \times 10^{9}$ & $6.8 \times 10^{8}$ & $9800$  & $17640$ & $8.2 \times 10^{6}$ & $9.8 \times 10^{5}$ \\
$\kappa(\mathbf{H}^T\mathbf{H})$  & $1$ & $8.5 \times 10^{5}$            & $8.4 \times 10^{5}$            & $1.1 \times 10^{9}$           & $1.3 \times 10^{8}$           & $7757.30$ & $12450.4$ & $2.0 \times 10^{6}$ \\
\end{tblr}\vspace{-0.6cm}
\end{table*}
\newpage
\textbf{Mixed PSFs invertibility:}
{Currently, various {DOEs} have been designed to solve specific tasks in hyperspectral imaging~\cite{jeon2019compact} or privacy preservation~\cite{arguello2022optics}. Consequently, there is a wide range of PSFs that can be analyzed to measure their invertibility in {Table \ref{tab:psfs}}. These include the unit impulse, Fresnel~\cite{yeh2010analysis}, Spiral~\cite{jeon2019compact}, and motion blur~\cite{dai2008motion}, where for the latter three, both a focused and a spread case are analyzed. In the {Table \ref{tab:psfs}}, it is possible to observe, for the cases where the PSF is focused and spread, both the condition number and the proposed metric accurately measure the invertibility of each matrix according to its similarity to the unit impulse, thus showing that the focused PSFs always have the highest invertibility. 
% The privacy PSF, and the spread cases of Fresnel and Spiral are the least invertible.
}

\begin{figure}[!t]
    \centering
    \includegraphics[width=0.92\linewidth]{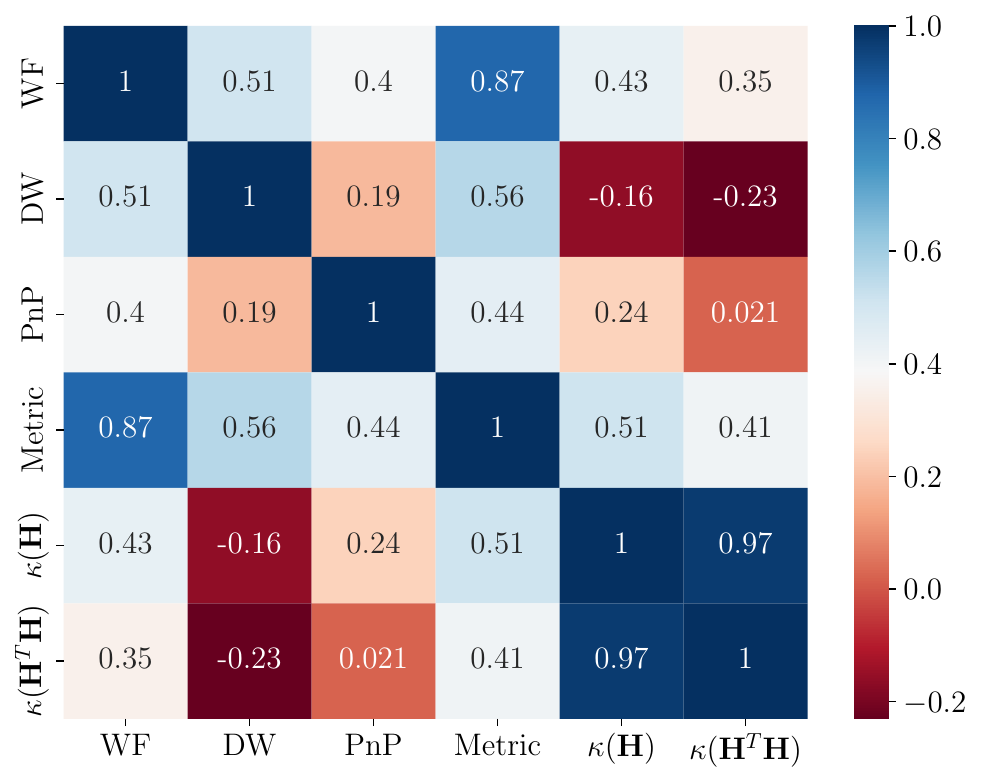}\vspace{-0.2cm}
    \caption{Correlation matrix between reconstruction performance in MSE, the proposed metric, and the condition number of the convolution matrix.}\vspace{-0.5cm}
    \label{fig:corr_psfs}
\end{figure}

\textbf{Correlation between invertibility and reconstruction for Mixed PSFs:}
% \romario{In \red{Fig. \ref{fig:corr_psfs}}, a study was conducted using different recovery approaches for ID, based on traditional, variational, and deep learning-based methods, such as the Wiener Filter~\cite{wiener1949extrapolation}, Plug-and-Play~\cite{chan2016plug}, and Deep Wiener~\cite{dong2020deep}.  Additionally, the method based on Deep Learning generally achieves the best reconstruction quality. Therefore, if one aims to leverage its potential, it is important to efficiently measure the suitability of the PSFs intended for deconvolution.}
{A correlation matrix was created to show the relationship between the proposed metric, the condition number of the convolution matrix, and the reconstruction performance for the ID problem. Three deconvolution approaches were used;
the Wiener filter, a variational method through a PnP algorithm \cite{chan2016plug}, using as a denoiser a {DNN and a DL-based} method inspired on the WF, the DW network~\cite{dong2020deep}. In the correlation matrix in Fig. \ref{fig:corr_psfs} it can be seen how the proposed metric has the highest correlation with all reconstruction methods. It is possible to conclude that the proposed metric is an effective method to measure the invertibility of a PSF in the {ID problem efficiently}.}
\begin{figure}[!t]
    \centering
    \includegraphics[width=0.92\linewidth]{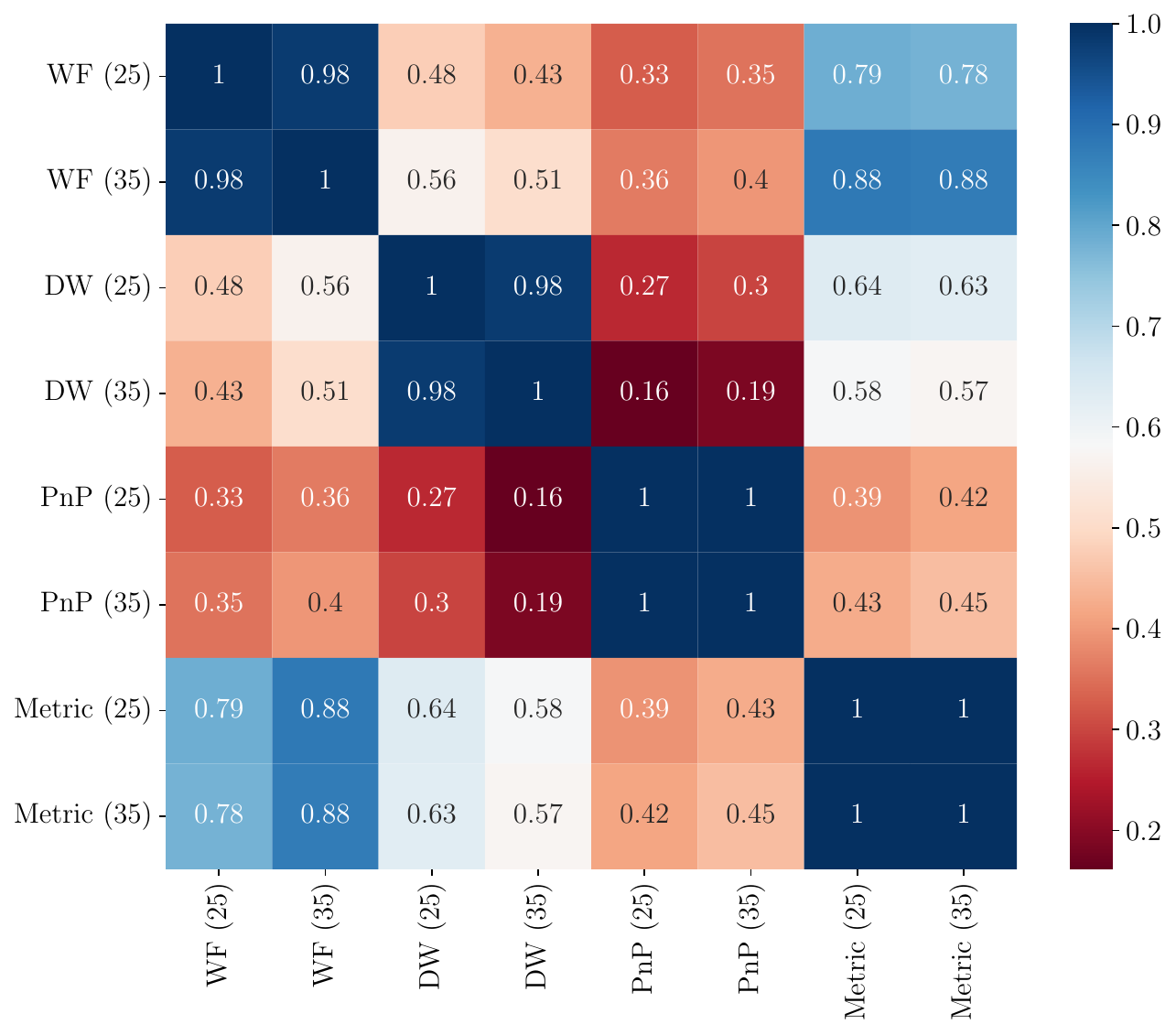}\vspace{-0.2cm}
    \caption{Correlation matrix for different noise levels in a data-driven scheme.}
    \label{fig:corr_noise}\vspace{-0.5cm}
\end{figure}

\textbf{Correlation with noise:}
{In this section, we focus on the relationship between the proposed metric and the deconvolution outcomes of various methods under the addition of white Gaussian noise at signal-to-noise ratio (SNR) levels of 25 and 35. This analysis is data-driven and was conducted using images from the BSDS500 dataset \cite{arbelaez2010contour}. The analysis, highlighted in Fig. \ref{fig:corr_noise}, demonstrates a significant correlation between the metric values and the performance of different reconstruction techniques. Specifically, the {WF and DW approaches show a high alignment with the {proposed metric and even the PnP performance has a notable correlation.} This} finding validates the utility of the proposed metric as a reliable indicator of the deconvolution method performance in noisy conditions.}

\begin{figure}[!t]
    \centering
    \includegraphics[width=0.95\linewidth]{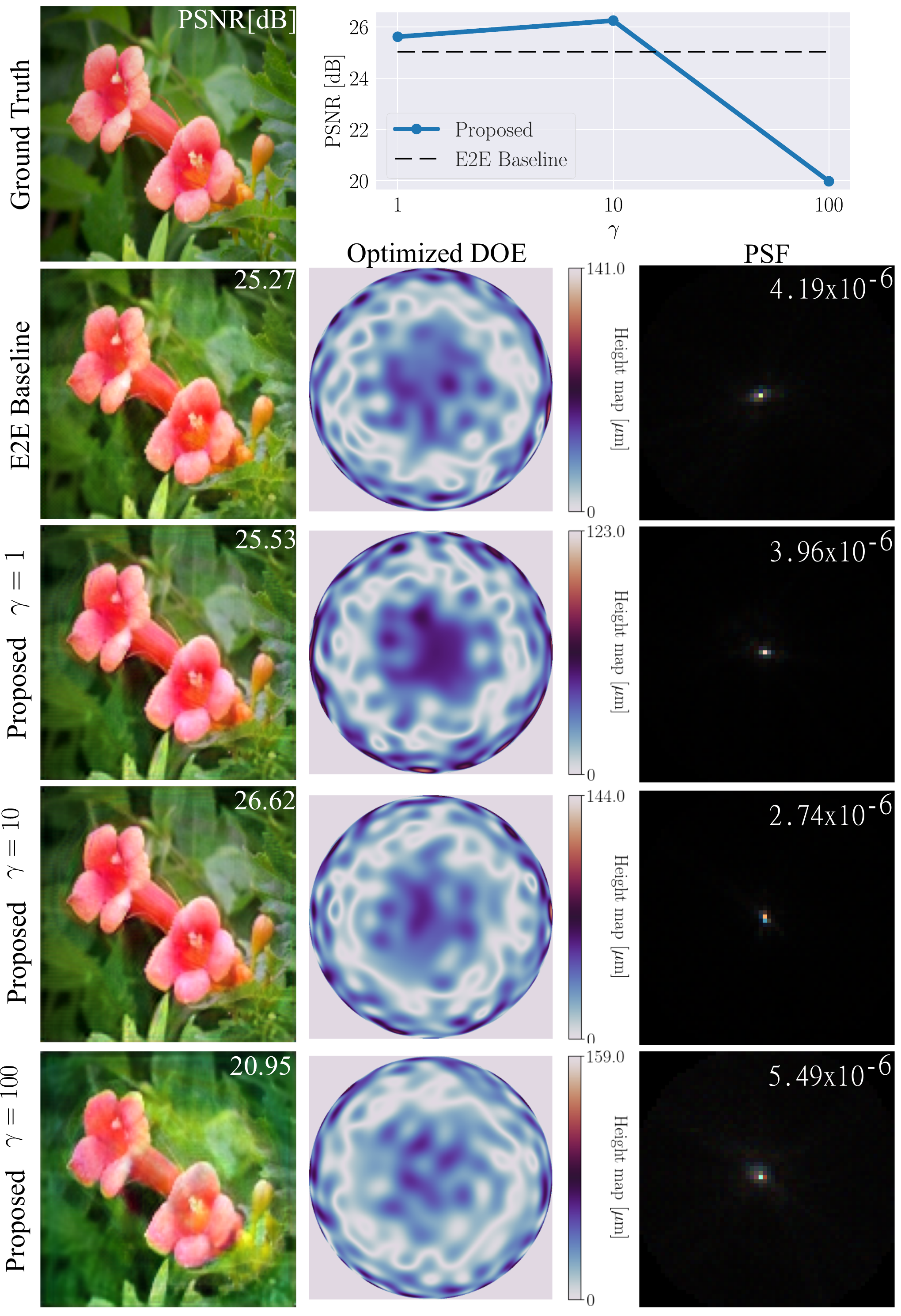}\vspace{-0.3cm}
    \caption{{Comparative results illustrating the impact of incorporating the proposed invertibility metric into the E2E optimization loss function for ID.}}\vspace{-0.6cm}
    \label{fig:e2e_results}
\end{figure}

\textbf{Invertibility metric as a regularizer in E2E optimization:}
Finally, in the pursuit of designing a DOE with an enhanced invertibility PSF, the proposed invertibility metric was incorporated into the loss function of an E2E optimization framework. Here we used a UNet architecture \cite{ronneberger2015u} as the recovery network $\mathcal{M}_\theta$. A total of $L=350$ Zernike polynomials were employed. This {approach aimed not only to optimize the DOE heightmap but also to ensure that the produced PSF facilitates} improved image recovery. 
% The inner optimization in \eqref{eq:inner} is solved with Adam optimizer \cite{kingma2014adam} with a learning rate of $1\times 10^{-3}$ for 500 epochs. Adam algorithm is employed to solve the outer loop using the same learning rate for 100 epochs. 
The inner optimization in \eqref{eq:inner} and the outer loop both use the Adam optimizer \cite{kingma2014adam} with a learning rate of $1\times 10^{-3}$, for 500 and 100 epochs, respectively.
\\
\\
To evaluate the effectiveness of incorporating the invertibility metric, experiments were conducted varying the regularization parameter $\gamma$, specifically testing for $\gamma = 1$, $\gamma = 10$, and $\gamma = 100$, which are shown in Fig. \ref{fig:e2e_results}. The results demonstrate a notable improvement in the quality of the DOE design, particularly for $\gamma = 1$ and $\gamma = 10$, with the latter achieving the most significant enhancement. A PSNR improvement of over 1 dB was observed when $\gamma = 10$, indicating the substantial impact of the proposed metric on the E2E optimization process.

\section{Conclusion}

An efficient differentiable nonlinear metric has been developed to assess the PSF invertibility in image deconvolution. This metric differs from traditional methods that rely on the analytical properties of the PSF by using a fully connected neural network to decrease computation time while maintaining effectiveness. As a result, it provides a quantifiable metric to evaluate the suitability of the PSF for deep learning-assisted recovery. The metric can be used as a regularizer in the design of diffractive optical elements requiring high invertibility. Future work could focus on other types of PSF such as spatially variant PSF which arises as an interesting scenario to employ the proposed approach.

\tiny{
\bibliographystyle{ieeetr}
\bibliography{biblio.bib}
}

\end{document}